\documentclass[%
 reprint,
 amsmath,amssymb,
 aps,
]{revtex4-2}

\usepackage{graphicx}
\usepackage{dcolumn}
\usepackage{bm}
\usepackage{xcolor}
\usepackage{soul}
\usepackage{float}
\usepackage{titlesec}
\usepackage{pgffor}    
\usepackage{float}     
\usepackage{orcidlink}

\titlespacing\subsection{0pt}{12pt plus 4pt minus 2pt}{0pt plus 2pt minus 2pt}

\begin{document}
\preprint{APS/123-QED}\

\title{Macroscopic Spin-Orbit Interaction through Strong-Field Pumping of Inhomogeneously Aligned Molecular Ensemble}


\author{Uriel Zanzuri}
    \email{urielzanzuri@mail.tau.ac.il}
 
\affiliation{%
 The Raymond and Beverly Sackler Faculty of Exact Sciences \\ Tel Aviv University, Tel Aviv 6997801, Israel
}%

\author{Sharly Fleischer~\orcidlink{0000-0003-0213-2165}}
\affiliation{Physical Chemistry Department, Raymond and Beverly Sackler Faculty of Exact Sciences, Tel Aviv University, Tel Aviv 6997801, Israel}

\author{Tamar Seideman~\orcidlink{0000-0002-0643-3742}}
\affiliation{Department of Chemistry, Northwestern University, Evanston, Illinois 60208, USA}

\author{Eldad Yahel}
\author{Amir Natan~\orcidlink{0000-0003-4517-5667}}
\email{amirnatan@tauex.tau.ac.il}
\author{Alon Bahabad~\orcidlink{0009-0003-7015-4611}}
\email{alonb@tauex.tau.ac.il}

\affiliation{
The School of Electrical Engineering, Tel Aviv University, Tel Aviv 6997801, Israel
}%

\date{\today}

\begin{abstract}
We study the strong-field interaction of a helical bi-chromatic pump with an anisotropic and inhomogeneous molecular system in the form of planar distribution of radially aligned molecular ensemble. This setting gives rise to macroscopic spin-orbit interaction where High Harmonic radiation is emitted while imbued with Orbital Angular Momentum (OAM) whose sign is directly dictated by the helicity of the pump field. We demonstrate this phenomenon in ensembles of  $H_2^+$ and $N_2$ molecules with Time-Dependent Density Functional Theory (TDDFT) simulations.

\end{abstract}

\maketitle

\emph{Introduction}. High Harmonic Generation (HHG) is an extreme nonlinear optical process in which an ionizing ultrashort pump field interacts with a medium. HHG was studied extensively within the gas phase since 1988 \cite{ferray1988multiple,ghimire2019high} with particular attention to interaction with aligned molecules \cite{itatani2005controlling,kanai2005quantum,ramakrishna2007information,mairesse2008high,marceau2020simultaneous}. These studies have revealed effects associated with molecular symmetry \cite{kanai2005quantum}, the dependence of HHG amplitude and polarization state on the molecular angle \cite{mairesse2008high} and more. The vast majority of these works considered effects explained at the single emitter level, which entails a homogeneous molecular ensemble  - that is -  the alignment of all molecules is the same everywhere within the plane of interaction with the HHG pump field. 
Here we consider the strong-field interaction of a pump field with a molecular ensemble that is both inhomogeneous and anisotropic  - the molecules are all aligned but, the alignment direction is a function of the position in space. This kind of symmetry is known to induce macroscopic spin-orbit coupling and was employed in various devices \cite{cardano2015spin,rubano2019q}, which essentially are an embodiment of the q-plate \cite{rubano2019q}, the first such device suggested and implemented using liquid crystal substrate. In a macroscopic spin-orbit coupling, also known as spin-orbit interaction \cite{cardano2015spin}, the spin angular momentum (SAM) of light, manifested by the polarization state of light, is coupled to the orbital angular momentum (OAM) of light. The latter is linked to the presence of a helical phase front in the field's amplitude of the form $exp(il\theta)$ where $\ell$ is known as the topological charge. Light impinging on a q-plate can acquire OAM (or change its value) in a way that is dependent on the value of its SAM. 
The use of a molecular q-plate was recently demonstrated in three key works.  Trawi et al. \cite{trawi2023molecular} successfully exploited a molecular q-plate to create a molecular quantum interface for storing and retrieving OAM. Xu et al. \cite{xu2025molecular} created a highly efficient, broadband molecular wave plate for generating ultrashort optical vortices across UV-IR wavelengths. Voisine et al.\cite{voisine2026molecular} have demonstrated that the vortex beam can be readout 'on demand' using rotational echo scheme. 
Here we extend the application of the molecular q-plate into the strong-field regime to manipulate HHG.
Explicitly, we show that when the molecular q-plate is pumped with a bi-chromatic helical pump consisting of circularly polarized fields, the emitted high harmonics carry OAM whose polarity depends on the SAM of the pump components.

Real-Time TDDFT (RT-TDDFT) \cite{marques2012fundamentals} is a powerful method for calculating the non-linear response of electrons within molecules to a strong laser field in the semi-classical approximation. RT-TDDFT was used in many works to calculate the response to a linear \cite{luppi2012computation,mack2013exchange,heslar2007high}, circular \cite{baer2003ionization,zhou2021high} and bi-chromatic circularly  polarized (BCCP) laser field \cite{koushki2018high}. In this work, we use RT-TDDFT to calculate the response of aligned $H_2^+$ and $N_2$ molecules to BCCP laser. Accounting for the emission from the entire radially aligned molecular ensemble, we find that the far field radiation carries OAM at selected harmonics, providing new means for generating OAM beams across various frequencies of interest.\\

\emph{Methodology}.
In our theoretical framework, we model the interaction of a gas layer composed of symmetric di-atomic homonuclear molecules radially aligned within a constrained spatial dimension. The molecular layer's thickness is small to mitigate phase mismatch phenomena within the medium, \cite{hareli2020phase}. De facto, the medium consists of a single layer of these molecules.

The molecules are pre-aligned via the response elicited by a strong laser pulse \cite{stapelfeldt2003colloquium, fleischer2012molecular, koch2019quantum, ohshima2010coherent}, where the alignment is primarily driven by the induced dipole interaction, where $\hat{V}=-\frac{1}{4}|E|^2  (\Delta \alpha \cos^2{\theta}\ + \alpha_{\perp})$. Here, $\Delta \alpha=\alpha_{\parallel}-\alpha_{\perp}$ is the anisotropic molecular polarizability, $E$ is the electric field applied on the molecule, and $\theta$ is the angle between the molecular axis and the field polarization axis.

The aligning beam is configured as a radial linear vector beam — a beam whose polarization state is linear at every point but aligns radially around the center of the beam. This specific polarization state is achievable, for example, by transmitting a uniformly linearly polarized beam through a q-plate characterized by a topological charge of \(q_p=\frac{1}{2}\) \cite{rubano2019q}.

Following the interaction with the aligning pulse, the molecules realign in a periodic manner known as 'rotational revivals' \cite{stapelfeldt2003colloquium, averbukh1989fractional, averbukh2001angular}, i.e. the aligned molecules can be further interrogated under field-free conditions by the HHG pump. The duration of each alignment event depends on the molecular moment of inertia, the temperature of the ensemble, as well as on the parameters of the alignment pulse (duration and intensity) \cite{stapelfeldt2003colloquium,leibscher2003molecular}, in the range of a few tens of fs for small diatomics and up to a few ps for large ones \cite{renard2003postpulse,rosca2002revival}. We therefore treat the 'aligned' molecules as static rotors during their interaction with the ultrashort HHG pump field (chosen as much shorter than the duration of alignment) for brevity.

The simulations were conducted at two distinct levels: microscopic and macroscopic. The microscopic simulations utilize RT-TDDFT through the Bayreuth version of the PARSEC package \cite{kronik2006parsec,mundt2007photoelectron, mundt2009real,hofmann2012kohn}. These simulations explore the dynamic response of aligned diatomic molecules subjected to a BCCP HHG pump field at varying relative alignment angles. 

\begin{figure}[H]
    \centering
    \includegraphics[scale=0.14]{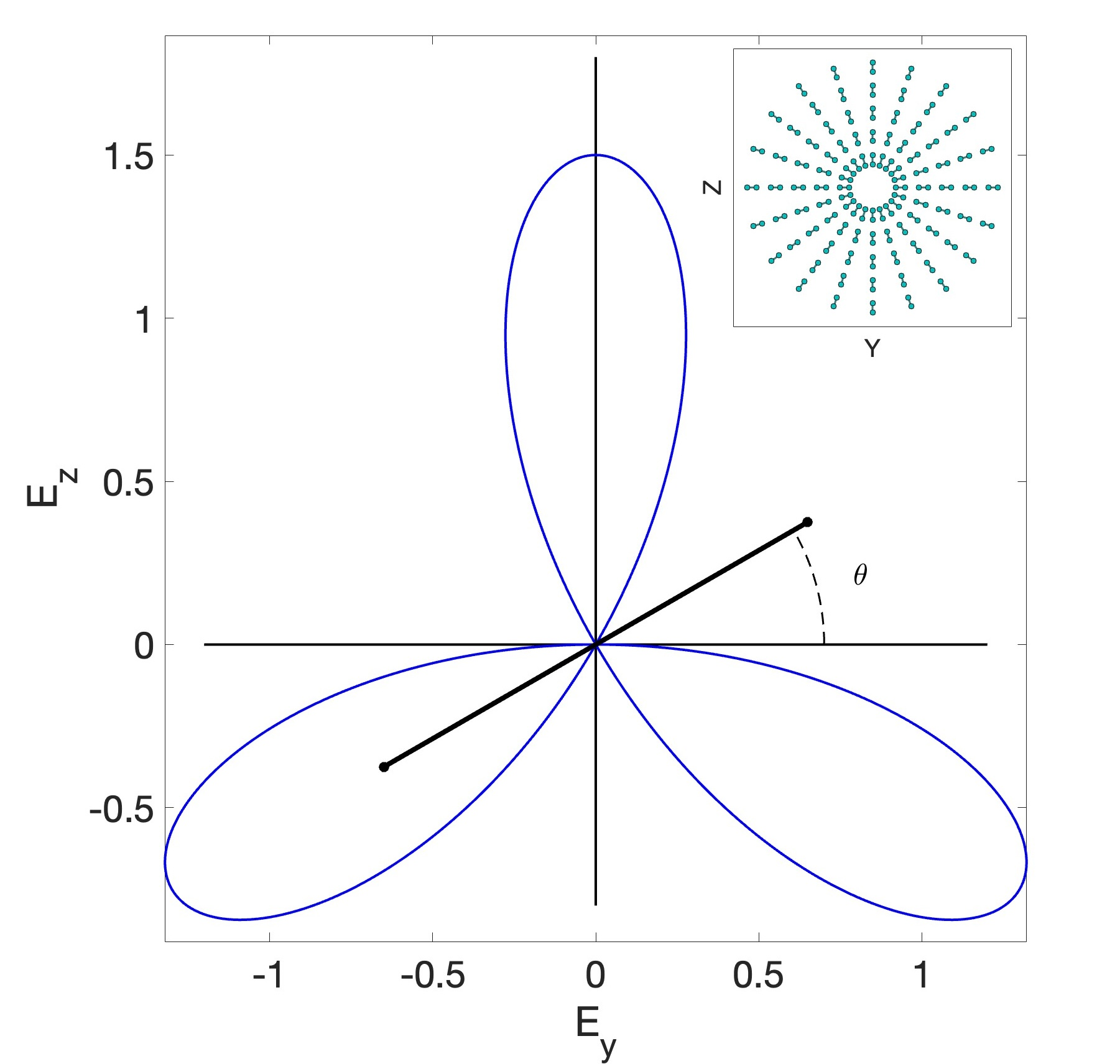} 
    \caption{\label{fig: orientation} A schematic cross-section of a bi-chromatic laser field interacting with a di-atomic molecule, oriented at an angle $\theta$ to the bichromatic field. The inset illustrates the molecular ensemble orientation. }
\end{figure}

The HHG pump field is configured as the composite of two counter-rotating circularly polarized field components, oscillating at the fundamental frequency $\omega_0$ and its second harmonic $2\omega_0$:

\begin{equation}
   \bold{E}_1(t)=\frac{1}{\sqrt{2}}E_0f(t)[\hat{\mathbf{e}}_y cos(\omega_0t)+\hat{\mathbf{e}}_z sin(\omega_0t)] \label{bi-circuar 1}
\end{equation}
\begin{equation}
   \bold{E}_2(t)=\frac{1}{\sqrt{2}}E_0f(t)[\hat{\mathbf{e}}_y cos(2\omega_0t)-\hat{\mathbf{e}}_z sin(2\omega_0t)] \label{bi-circuar 2}
\end{equation}

where $f(t)$ represents the envelope of the pulse. 
Here the overall field is a right-handed BCCP field for which the fundamental is Right-Handed Circularly polarized (RHC) and the second-harmonic is Left-Handed Circularly polarized (LHC).
The BCCP field, known for its efficiency in generating circularly polarized HHG from atoms \cite{fleischer2014spin}, possesses a helical structure. The helicity of this field can be altered by exchanging between the circular polarization helicity of its two components. The trajectory described by the tip of this field in space, along with a schematic of an aligned molecule interacting with it, is depicted in Fig. \ref{fig: orientation}.

For the TDDFT simulations, the evolution of Kohn-Sham (KS) orbitals is approximated using a fourth-order Taylor expansion combined with a predictor-corrector method \cite{mundt2009real}. These simulations provide the dynamics of the orbitals under the influence of the laser field.

We focus on the induced dipole components along the $y$ and $z$ axes. From the Fourier transform of the time-dependent dipoles, we derive the harmonics spectrum in both the $y$ and $z$ directions. A custom script has been developed to locate and record the complex value of each harmonic peak. The magnitude of the peak represents the absolute value, indicative of the point source amplitude, whereas the angle provides the phase of this source. The mathematical representation of the time-dependent dipole and its Fourier transformation is given by:

\begin{equation}
    \vec{d}(t) = \int d\vec{r} \vec{r} \rho(\vec{r},t), \quad
    \vec{d}(\omega) = \frac{1}{T} \int_0^T dt e^{-i\omega t} \vec{d}(t) \label{dipole_fft}
\end{equation}

where $\rho(\vec{r}, t)$ denotes the electron density at position $\vec{r}$ and time $t$, $\vec{d}(t)$ represents the time-dependent dipole, and $\vec{d}(\omega)$ is its Fourier transform in the frequency domain.

We simulated the interaction for both $H_2^+$ and $N_2$ molecules. Technical information about the simulation parameters is given in section II of the supplementary material (SM) section III. For $H_2^+$, which contains only a single electron, the simulation employs a pure Time-Dependent Schrödinger Equation (TDSE) approach, capturing the dynamics of the electron under the applied field. $N_2$, being a multi-electron system, necessitates the use of Time-Dependent Density Functional Theory (TDDFT) incorporating the asymptotically corrected van Leeuwen and Baerends (LB) exchange functional \cite{van1994exchange}, which previously was shown to predict a realistic ionization potential for $N_2$ \cite{chu2004role}.

The macroscopic part of the simulation uses the outcomes of the microscopic analysis as an input for propagating the dipole radiation emitted from the molecules in the ensemble to the far field. The following results highlight how the macroscopic response mediates a spin-orbit coupling. Specifically, the OAM state of the far-field radiation is shown to depend on the helicity of the pump, which is directly dictated by the SAM of its two components (fundamental and second harmonic).

\emph{Results}.
We begin by presenting the TDSE microscopic simulation results for the \(H_2^+\) molecule, which serves as an ideal numerical system for studying the generated dipole moment as a function of its alignment relative to the bi-chromatic electric field orientation, as illustrated in Fig. \ref{fig: orientation}.

\begin{figure}[H]
    \centering
    \includegraphics[scale=0.095]{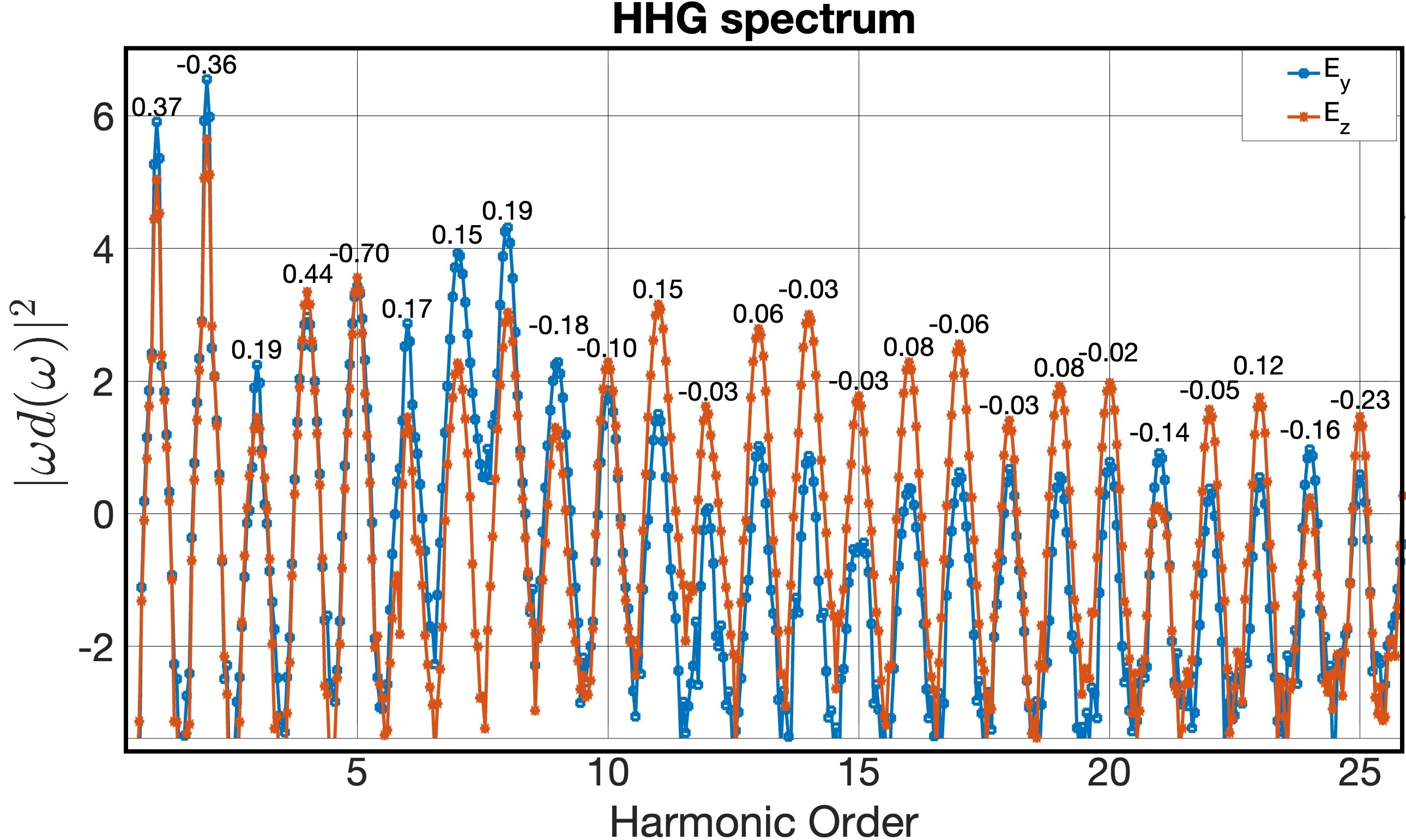}
    \caption{Harmonic spectra of \(H_2^+\) for $\theta=0^\circ$ (as defined in Fig. \ref{fig: orientation}) under a bi-chromatic laser field. The ellipticity for each harmonic order is shown above the peaks, the sign of which represents the handedness of the field.}
    
    \label{fig:harmonics_0deg}
\end{figure}

A representative case is displayed in Fig.\ref{fig:harmonics_0deg}, where the molecule is aligned at a relative angle of \(\theta = 0^\circ\) 
to the bi-chromatic electric field. This example reveals that harmonics of even order, which are forbidden in linear polarization, are generated under the counter-rotating bi-chromatic field \cite{neufeld2019floquet}.

\begin{figure*}[ht]  
    \centering
    \includegraphics[scale=0.17]{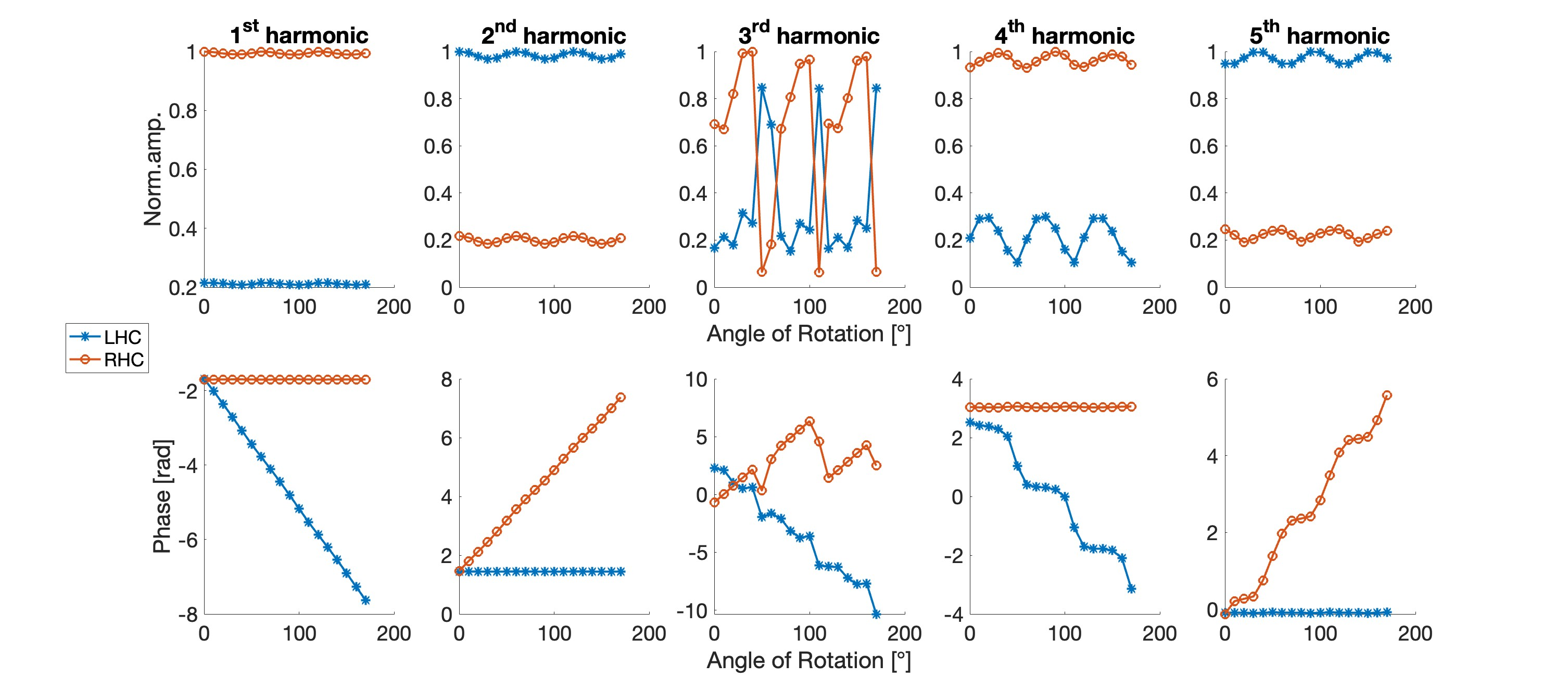}
    \caption{Normalized amplitude (divided by the maximal value) and phase of the 1$^\mathrm{st}$-5$^\mathrm{th}$ harmonics of \(N_2\) subjected to a counter-rotating bi-chromatic laser driving field for LHC (blue) and RHC (red) polarization states.}
    \label{fig:amp_phase_harmonics}
\end{figure*}

Our main interest here is in systems that can be readily tested in the laboratory. As such, the $N_2$ molecule, which has served as a case study in most experiments on HHG from molecules, is a useful example explored in the remainder of the paper (additional results of H$_2^+$ are provided in sections IV and V of the SM). The HHG spectrum for the case of  $N_2$ molecule is given in section IV of the SM. The amplitude and phase of the $1^{st}$ to the $5^{th}$ harmonics of $N_2$ field decomposed to the left and right circular polarization states, as a function of the alignment angle $\theta$, are shown in Fig. \ref{fig:amp_phase_harmonics}. Due to the inversion symmetry of the molecules along their molecular axis, the response of these harmonics exhibits periodic behavior with a period of 180 degrees. Notably, several harmonic orders display an almost linear dependence of the phase on the alignment angle. This characteristic is crucial for achieving integer values of OAM in the far-field radiation generated by the molecular ensemble. {To check the possible effect of field intensity on the phase and amplitude response, we compare, in the SM (section IV.E), the response of $\mathrm{N_2}$ at field intensities of $1\times 10^{14}$ and $3\times 10^{14}\ \mathrm{W/cm^2}$, and of $\mathrm{H_2^+}$ at field intensities of $3\times 10^{13}$, $1\times 10^{14}$, and $3\times 10^{14}\ \mathrm{W/cm^2}$. We show that the phase response is only weakly affected by the field intensity.}

To corroborate the simulations, we developed a mathematical model describing the linear response of a diatomic molecule to a BCCP Field (see SM). We observed a good agreement of the results of this analytical theoretical calculation with the simulation results. Moreover, we observed that the fundamental and second harmonic show a value much smaller than one for the ellipticity, which is attributed to the anisotropy of the molecular polarizability tensor. This is further analyzed in the SM.   

Next we present the results of far-field radiation from the macroscopic molecular ensemble. The dipole radiation from each molecule in a radially aligned ensemble is used to compute the overall far-field radiation. The field for each harmonic order $q$ was calculated at a distance of $2 \times \lambda_q$, where $\lambda_q$ is the wavelength of the $q^{th}$ harmonic. The molecules were arranged on five concentric circles with equal spacing between each circumference. In this configuration, the outermost radius was of the order of the harmonic wavelength.  The configuration of the molecular ensemble is depicted in Fig. \ref{fig: orientation}.

The molecular ensemble was subjected to three different field configurations: a linear bi-chromatic field, a clockwise BCCP, and a counterclockwise BCCP field. These latter two fields were created using circularly polarized components, with the helicity switched between the two cases to explore the effects of field helicity on the far-field radiation patterns.

We then analyzed the far-field radiation patterns for several high harmonic frequencies. The fields were decomposed into a finite basis of Laguerre-Gauss modes, which are useful in evaluating the OAM possessed by a given field. This decomposition involved projecting the field onto the first 189 modes, spanning azimuthal indices \(\ell = -4\) to \(\ell = 4\) and radial indices \(p = 0\) to \(p = 20\).

The mathematical form of these modes, in cylindrical coordinates $(r, \phi, z)$ is:

\begin{align}
\begin{split}
\psi_{lp}(r, \phi, z) = & \sqrt{\frac{2p!}{\pi(|\ell|+p)!}} \cdot \frac{1}{w(z)} \cdot \left(\frac{\sqrt{2}r}{w(z)}\right)^{|\ell|} \\
& \cdot \exp\left(-\frac{r^2}{w^2(z)}\right) \cdot L^{|\ell|}_p\left(\frac{2r^2}{w^2(z)}\right) \cdot \exp(i\ell\phi) \\
& \cdot \exp\left(-i\frac{kr^2}{2R(z)}\right) \cdot \exp(i(2p+|\ell|+1)\xi(z)),
\label{Laguerre-Gauss}
\end{split}
\end{align}

where $L^{|\ell|}_p$ is the generalized Laguerre polynomial, $w(z)$ is the beam width, $R(z)$ is the radius of curvature, and $\xi(z)$ is the Gouy phase shift \cite{siegman1986lasers}.

From the analysis of the far-field patterns, we observed that no OAM was generated under linear polarization (as an example refer to the response at the fundamental harmonic presented in section IV of the SM). This outcome is as expected because a linearly polarized field does not exhibit the necessary symmetry in its phase and amplitude distributions to impart OAM. In essence, the dipole response induced by a linearly polarized bi-chromatic field tends to maintain symmetry that does not support the generation of OAM.

Conversely, clear OAM patterns emerged when we used BCCP fields endowed with SAM. The generated OAM changes sign with the switching of the field helicity, which exemplifies the spin-orbit interaction. This phenomenon associates the angular momentum states of the photons in the driving laser field with the resultant harmonic. This shows that the system as a whole can be interpreted as a coherently controlled molecular q-plate.

\begin{figure*}[ht]  
    \centering
    \includegraphics[scale=0.17]{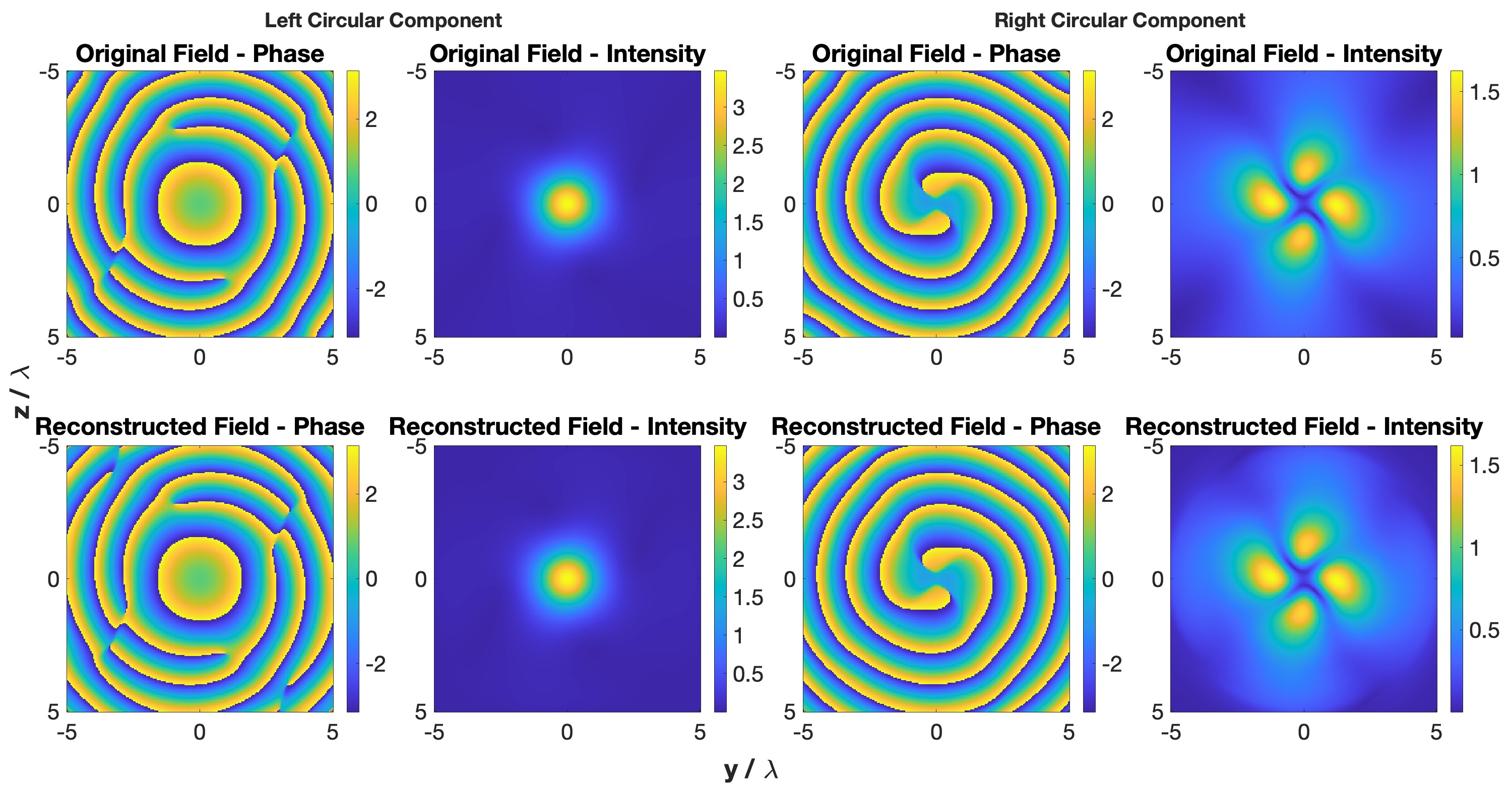}
    \caption{Far-field pattern of the \(5^{\mathrm{th}}\) harmonic of \(N_2\) driven by the counter-rotating bi-chromatic circularly polarized (BCCP) pump field, resolved into its left and right-circular components. For each component the upper row is the field obtained from the simulation and the lower row its reconstruction from the first \(189\) Laguerre-Gauss modes (\(\ell=-4\ldots4\), \(p=0\ldots20\)). Within each row the left panel shows the phase and the right panel the intensity. The field is evaluated at a distance \(x=2\lambda_q\) on a screen of half-width \(5\lambda_q\) in units of the harmonic wavelength \(\lambda_q\). The right-circular component carries the orbital angular momentum. Reversing the pump helicity interchanges these roles (see section V.B and V.C of the SM).}
    \label{fig:N2_el_er_field}
\end{figure*}

\begin{figure*}[ht]  
    \centering
    \includegraphics[scale=0.23]{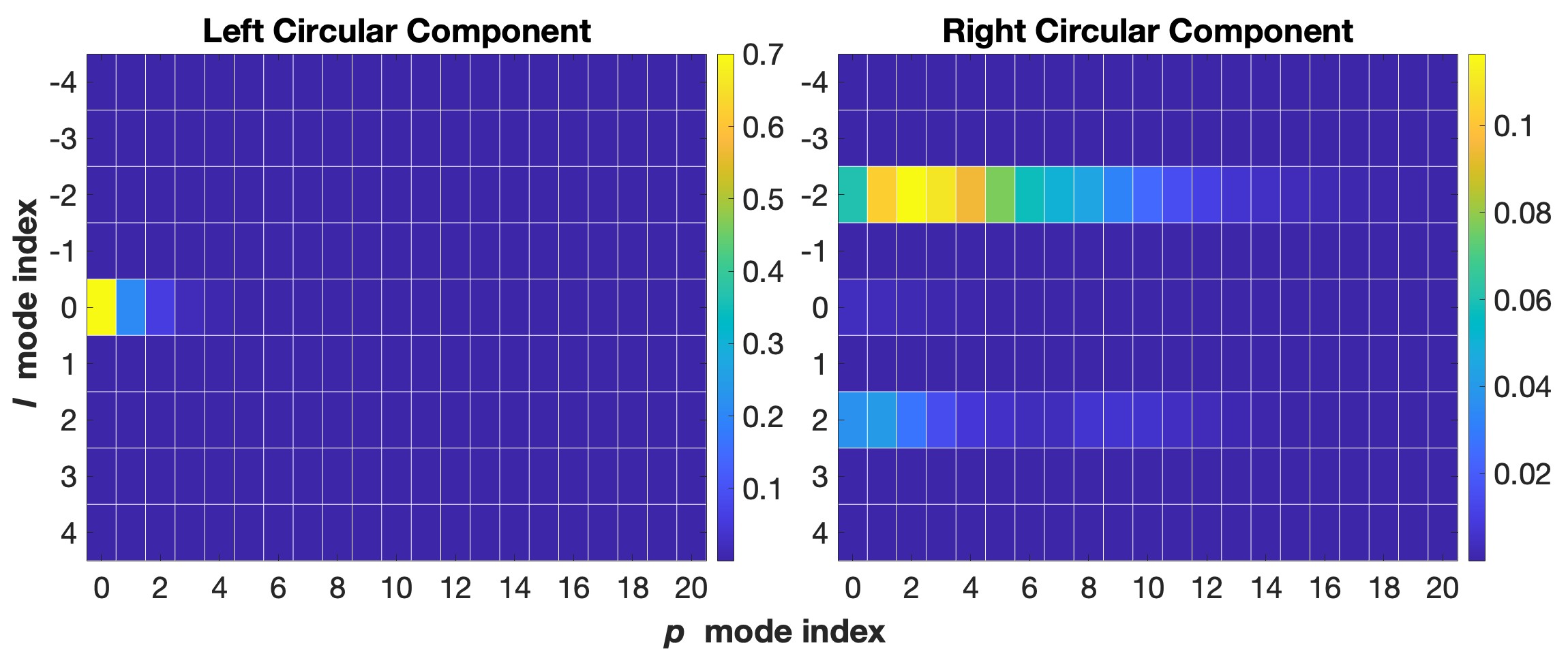}
    \caption{Laguerre-Gauss decomposition of the \(5^{\mathrm{th}}\) harmonic far field of \(N_2\) under the BCCP pump field, shown for the left and right-circular components (left and right panels). The radial index \(p\) runs along the columns and the azimuthal index \(\ell\) along the rows, the color encodes the relative weight of each mode. The right-circular component projects predominantly onto \(\ell=-2\), with a smaller \(\ell=+2\) contribution, identifying the emitted OAM, while the left-circular component is concentrated at \(\ell=0\).}
    \label{fig:N2_el_er_mat}
\end{figure*}

For illustrative purposes, Fig.\ref{fig:N2_el_er_field} displays the left and right circularly polarized components of the 5th harmonic in the far field for $N_2$ molecules. These figures include both amplitude and phase details for the fields and their reconstruction from 189 Laguerre-Gauss modes, spanning azimuthal indices \(\ell = -4\) to \(\ell = 4\) and radial indices \(p = 0\) to \(p = 20\). The relative contribution of each mode is displayed as a heatmap in Fig. \ref{fig:N2_el_er_mat}. These results can be explained using an intuitive approach based on selection-rules which is given in detail in the SM. The main point of this explanation is to first treat the molecules as point-source while using SAM selection rules to determine the dominant polarization state of the emitted radiation. However, as the molecules are anisotropic, an additional polarization state would be excited. It is assumed that this additional polarization state is phase-locked to the orientation of the molecule.  These assumptions, when taken for all molecules in the macroscopically ordered ensemble of the molecular q-plate, yield the correct SAM and OAM states of the emitted radiation. {Fig.~\ref{fig:N2_el_er_mat} shows a large projection onto $\ell=-2$ and a much smaller projection onto $\ell=2$ for the right circular component. Some $\ell=2$ character could arise from the beam's radial expansion relative to the source; in the SM (section VI.B), we analyze the effect of the size of the molecular source and show that, for a ring of molecules with a larger radius, we obtain only the $\ell=-2$ term.}

\emph{Summary}.
We investigated macroscopic spin-orbit interaction in a radially aligned molecular gas ensemble driven by a strong helical bi-chromatic pump field. Using real-time Time-Dependent Density Functional Theory (RT-TDDFT) simulations, we demonstrated that high harmonic radiation collectively emitted from judiciously patterned ensemble of molecular orientations carries OAM whose sign depends on the pump field's helicity. This effect, observed in $H_2^+$ and $N_2$ molecules, resembles the behavior of a q-plate, enabling coherent control of the OAM of the HHG radiated beams. Our results highlight the potential for structured light generation and macroscopic spin-orbit coupling in strong-field physics and open new avenues for ultrafast light-matter interaction.

\clearpage
%

\onecolumngrid

\newpage
\appendix

\end{document}